\documentclass[%
 preprint,
 amsmath,amssymb,
 aps,
prb,
]{revtex4-2}

\usepackage{graphicx}
\usepackage{dcolumn}
\usepackage{bm}
\usepackage{soul,color}

\begin{document}

\title{Spectral heat flux redistribution upon interfacial transmission} 
\author{Haoran Cui}%
\author{Theodore Maranets}
\author{Tengfei Ma}
\author{Yan Wang}
 \email{yanwang@unr.edu}
\affiliation{Department of Mechanical Engineering, University of Nevada, Reno, Reno, NV 89557}%

\date{\today}

\begin{abstract}
In nonmetallic crystals, heat is transported by phonons of different frequencies, each contributing differently to the overall heat flux spectrum. In this study, we demonstrate a significant redistribution of heat flux among phonon frequencies when phonons transmit across the interface between dissimilar solids. This redistribution arises from the natural tendency of phononic heat to re-establish the equilibrium distribution characteristic of the material through which it propagates. Remarkably, while the heat flux spectra of dissimilar solids are typically distinct in their bulk forms, they can become nearly identical in superlattices or sandwich structures where the layer thicknesses are smaller than the phonon mean free paths. This phenomenon reflects that the redistribution of heat among phonon frequencies to the equilibrium distribution does not occur instantaneously at the interface, rather it develops over some time and distance. 
\end{abstract}

\maketitle
\section{Introduction}
Heat transfer between nonmetallic crystals, particularly semiconductors used in electronic and photonic devices, is primarily facilitated by phonons. Even when two dissimilar solids are in atomic-level contact, thermal resistance at the interface ($R_{\text{int}}$), known as Kapitza resistance, can arise. As modern electronic and photonic devices shrink to the nanometer scale, $R_{\text{int}}$ becomes more dominant compared to the bulk thermal resistance within each component \cite{cahill2014nanoscale}. Therefore, it is crucial to develop a rigorous understanding of interfacial thermal transport.

Early models assumed elastic transmission of phonons, meaning heat is carried by phonons of the same frequency ($\omega$) across the interface, though these models differed in their treatment of transmission behaviors \cite{swartz1989thermal,chen2022interfacial}. In recent decades, more sophisticated approaches, including molecular dynamics and atomistic Green's functions supported by experimental evidence, have unveiled various intriguing details of interfacial phonon transport. Notably, inelastic phonon transmission, where heat is carried by phonons of different frequencies upon crossing the interface, has been found to reduce $R_{\text{int}}$ at higher temperatures \cite{hopkins2009multiple,saaskilahti2014role}. Additionally, phonon nonequilibrium  near interfaces, typically manifested as different phonon modes having different temperatures \cite{zhou2013phonon,ma2022ex}, has been shown to increase $R_{\text{int}}$, similar to how electron-phonon nonequilibrium increases the $R_{\text{int}}$ of metal-nonmetal interfaces \cite{majumdar2004role,wang2012two,wang2016effect,lu2019esee}. More recent research on multilayered structures has revealed fascinating behaviors, where interference, localization, and mode conversion can either increase, reduce, or even eliminate the effective $R_{\text{int}}$ of their interfaces \cite{luckyanova2012coherent,wang2014decomposition,luckyanova2018phonon,maranets2024influence,maranets2024prominent}.

Several essential aspects of interfacial thermal transport remain largely unaddressed. Notably, how does phonon nonequilibrium near the interface affect the heat flow carried by different modes, i.e., the heat flux spectrum $Q(\omega)$? What is the characteristic time or distance ($\delta_{t}$) needed for phonon heat fluxes to restore their thermal equilibrium distribution among phonon modes after interface transmission? Moreover, in scenarios involving two or more interfaces, such as a superlattice of two materials, what occurs if the spacing between interfaces is less than $\delta_{t}$, preventing phonon heat fluxes from fully thermalizing into their equilibrium distribution before encountering another interface? Answering these questions will not only improve our fundamental understanding of phonon transport across single and multiple interfaces but also pave the way for developing novel strategies to better control thermal transport in modern devices and advanced materials.

In this work, we address the above questions by rigorously analyzing the heat flux spectra along the heat flow direction for heterostructures containing single, double, and multiple interfaces, as discussed in the following section.

\section{Methodology}
\subsection{Model system}
We conduct nonequilibrium molecular dynamics (NEMD) simulations of Lennard-Jones (LJ) conceptual crystals using the LAMMPS package \cite{thompson2022lammps}, similar to our previous studies \cite{wang2014decomposition,chakraborty2020complex}. The potential well depth of the LJ potential is set as $\epsilon$ = 0.1664 eV, and the zero-crossing distance $\sigma$ is set as 3.4 \AA, with a cutoff radius of 2.5 $\sigma$. The use of $\epsilon$ = 0.1664 eV, which is sixteen times the value for solid argon, represents a material with much stronger bonding to mimic covalently bonded semiconductors. The two materials in the heterostructures are modeled with the same LJ parameters but with different atomic masses, 40 g/mol and 90 g/mol, are referred to as m40 and m90, respectively.

The models in our NEMD simulations have a cross-sectional area of 6 UCs $\times$ 6 UCs ($y$-$z$ plane), where 1 UC = 5.27 Å. The length of the device and each heat bath along the heat flow direction ($x$) is 1024 UCs and 512 UCs, respectively. Periodic boundary conditions are applied in all three dimensions. At the beginning of the simulation, we relax the structure through two stages of zero-pressure isothermal-isobaric (NPT) integrations. In the first NPT stage, the temperature increases from 5 K to the target temperature for 1 ns with a time step size of 1 fs. Subsequently, in the second stage, we maintain the system temperature at the target temperature over 2 ns. After the relaxation, an approximate 2 UC layer of atoms at both ends of the structure in the $x$ direction are frozen as the fixed boundary condition. The simulation is then switched to plain integration for 20 ns to reach a steady state, followed by another 10 ns to output the heat flux information. 
\subsection{Phonon heat flux spectrum}
To compute the phonon heat flux spectrum $Q(\omega)$, we monitor the interactions between atoms in the 10.54-Å thick region (2 UCs wide) on the left of an interface (actual or imaginary) and the 10.54-Å thick region on its right, denoted as $\tilde{L}$ and $\tilde{R}$, respectively. Then, $Q(\omega)$ is calculated as \cite{saaskilahti2014role,Cui20242Dmaterial}:
\begin{equation}
Q(\omega)=\sum_{i\in\tilde{L}}\sum_{j\in\tilde{R}} \left( -\frac{2}{t_{\text{simu}}\omega}\sum_{\alpha,\beta}\text{Im}\left\langle\hat{v}_{i}^\alpha(\omega)^*K_{ij}^{\alpha\beta}\hat{v}_j^\beta(\omega)\right\rangle \right),
\label{eqn:spectral_q}
\end{equation}
where $\hat{v}$ is the Fourier transform of atomic velocity, $t_{\text{simu}}$ is the time duration for the velocity data, and $K_{ij}^{\alpha\beta}$ is the force constant matrix. The subscripts $i$ and $j$ are the atom indices in the $\tilde{L}$ and $\tilde{R}$ regions, while the superscripts $\alpha$ and $\beta$ are Cartesian coordinates. $^*$ is the complex conjugate operator.

\subsection{Phonon Mean Free Path} 
The phonon mean free path (MFP) can be calculated as \cite{latour2017distinguishing}
\begin{equation}
\Lambda(\boldsymbol{k})=v_{g}(\nu,\boldsymbol{k})\times \tau_{\boldsymbol{k},\nu},
\label{eqn:MFP}
\end{equation}
where $\boldsymbol{k}$ is the wavevector, $\nu$ is the phonon polarization, $v_{g}(\nu,\boldsymbol{k})$ represents the group velocity of the phonon mode $(\nu,\boldsymbol{k})$, and $\tau_{\boldsymbol{k},\nu}$ is the phonon lifetime. The phonon group velocity can be calculated from the phonon dispersion relation obtained using the ALAMODE package \cite{TadanoPRB2015} based on the lattice dynamics theory. To obtain the phonon lifetime at 30 K, we conduct spectral energy density (SED) analysis \cite{Thomas2010PRB}, which quantifies the kinetic energy of phonon modes, widely applied to evaluate the phonon properties in various types of structures \cite{Thomas2010PRB,Cui2024PRB,Iyyappa2024CMS}.  The SED of different phonon modes can be evaluated by projecting the positions of the atoms in a crystal onto the time-domain normal mode coordinates, $q_{\boldsymbol{k},\nu}(t)$, which is defined as 
\begin{equation}
q_{\boldsymbol{k},\nu}(t) = \sum_{\alpha}^{3}\sum_{b}^{n}\sum_{l}^{N_c} \sqrt{\frac{m_b}{N_c}}u_{\alpha}^{l,b}(t)e_{b,\alpha}^{\boldsymbol{k},\nu*}exp[i\boldsymbol{k}\cdot \boldsymbol{r}_{0}^{l}],
    \label{eqn:normal_mode}
\end{equation}
where $u_{\alpha}^{l,b}(t)$ represents the $\alpha th$ component of the displacement of $bth$ basis atom in the $lth$ unit cell, $m_{b}$ is atomic mass, $N_{c}$ is the total number of primitive unit cells of the entire system, $t$ is time, and $\boldsymbol{r}_0$ is the equilibrium position of each unit cell. Subsequently, the SED is calculated through Fourier transform ($\mathcal{F}[\cdot]$) of the time derivative of $q_{\boldsymbol{k},\nu}(t)$ as
\begin{equation}
\Phi_{\boldsymbol{k},\nu}(\omega)=| \mathcal{F}[{\dot q_{\boldsymbol{k},\nu} (t)}]|^{2}=\frac{C_{\boldsymbol{k},\nu}}{(\omega-\omega_{\boldsymbol{k},\nu}^{A})^2+(\tau_{\boldsymbol{k},\nu}^{-1})^2/4},
\label{eqn:SED}
\end{equation}
where $\Phi_{\boldsymbol{k},\nu}(\omega)$ is the SED, $\omega$ is the angular frequency, and $C$ is a constant. The transformed data is fitted with the Lorentzian function to obtain the peak position $\omega_{\boldsymbol{k},\nu}^A$ and the full width at half maximum $\tau_{\boldsymbol{k},\nu}^{-1} $. Consequently, the phonon MFP can be calculated by substituting the phonon lifetime obtained from the Lorentzian function and the phonon group velocity obtained from the lattice dynamics into Eq.~\ref{eqn:MFP}.

\section{Results and Discussions}
\subsection{Single interface}

\begin{figure}
\centering 
\includegraphics[width=1\textwidth]{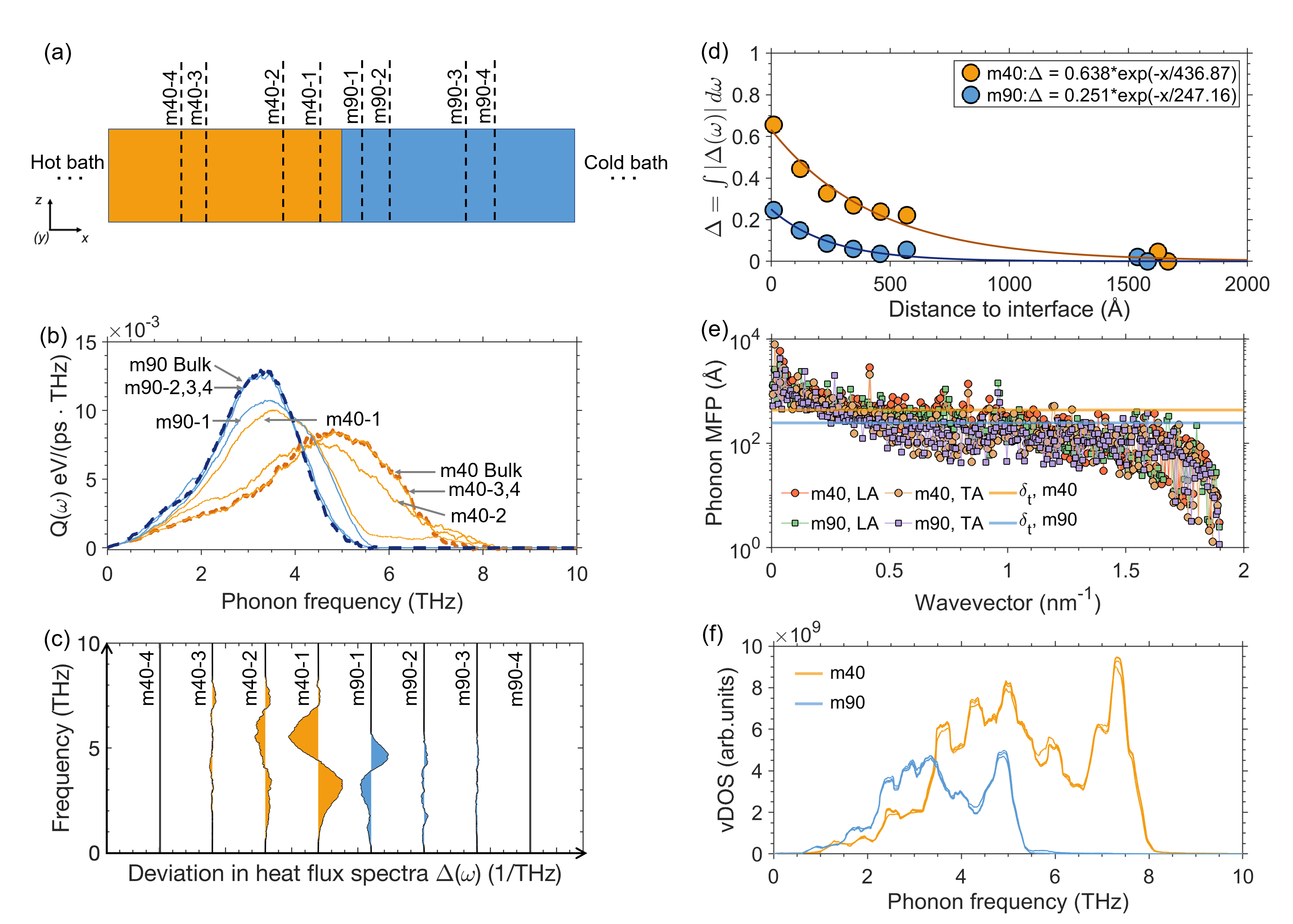}
\caption{(a) Schematic of a single interface structure composed of m40 and m90 conceptual Lennard-Jones crystals, containing 1024 UCs for each m40 or m90 part. The dashed lines indicate the positions where the heat flux spectra $Q(\omega)$ are extracted. (b) $Q(\omega)$ curves at different positions: gold curves represent locations within the m40 material, while blue curves correspond to those within the m90 material. Additionally,  the solid curves refer to the $Q(\omega)$ at different positions, while the thick dashed curves represent the $Q(\omega)$ of bulk m40 and m90 materials.  (c) Normalized deviation $\Delta(\omega)$ of the local heat flux spectrum from the reference spectrum based on Eq.~\ref{eqn:delta_omega}. (d) Variation of $\int |\Delta(\omega)| \, d\omega$ with distance to interface based on Eq.~\ref{eqn:delta}. Specifically, to enhance the accuracy of exponential fitting curves, we add 4 more positions within each m40 or m90 material, resulting in a total of 8 data points for each material. (e) Comparison of the phonon MFP for both LA and TA branches in m40 and m90. The gold and blue horizontal lines indicate the value of $\delta_{t}$ for m40 and m90. (f) Phonon vDOS at the corresponding locations, highlighting that the vDOS of the same material (m40 or m90) remains consistent, regardless of position, which differs from the behavior of $Q(\omega)$.}
\label{fig:SI}
\end{figure}

First, we investigate the evolution of the heat flux spectra $Q(\omega)$ in a single-interface system composed of m40 and m90 materials. As schematically illustrated in Fig.~\ref{fig:SI}a, we select eight distinct locations to extract $Q(\omega)$: four on the m40 side, positioned at distances of 11~\AA\ (m40-1), 570~\AA\ (m40-2), 1,620~\AA\ (m40-3), and 1,670~\AA\ (m40-4) from the m40-m90 interface. Symmetrically, four locations are selected on the m90 side, labeled as m90-1, m90-2, m90-3, and m90-4.

Figure~\ref{fig:SI}b presents the $Q(\omega)$ extracted at these locations. Notably, the $Q(\omega)$ at m40-2, m40-3, and m40-4 closely matches the $Q(\omega)$ of bulk m40. Similarly, the heat flux spectra at m90-2, m90-3, and m90-4 resemble the $Q(\omega)$ of bulk m90. These observations are expected, as m40/90-2/3/4 are sufficiently distant from the interface, allowing phonons to behave as though they are within the bulk material.

A striking observation is that $Q(\omega)$ at m40-1 and m90-1, which are located adjacent to the interface, are nearly identical, yet they differ significantly from their respective bulk counterparts, as shown in Fig.~\ref{fig:SI}b. This suggests that heat is predominantly carried by phonons of the same frequency as they transmit across the interface (i.e., elastic transmission). Away from the interface, anharmonic phonon scattering redistributes the heat flux among various phonon modes.

The evolution of $Q(\omega)$ along the direction of heat flow is further elucidated by defining the normalized deviation of the local $Q(\omega)$ from the bulk $Q(\omega)$ of the corresponding material (m40 or m90) as:
\begin{equation}
\Delta(\omega) = \frac{Q(\omega) - Q_{\text{ref}}(\omega)}{\int_{0}^{\omega} Q_{\text{ref}}(\omega) \, d\omega},
\label{eqn:delta_omega}
\end{equation}
where $Q_{\text{ref}}$ represents the $Q(\omega)$ of the corresponding bulk material (m40 or m90).

As shown in Fig.~\ref{fig:SI}c, the deviation is minimal far from the interface and becomes most pronounced near the interface (at m40-1 and m90-1). Specifically, at m40-1, the low-frequency range (0–4.4 THz) of $\Delta(\omega)$ is positive, while the high-frequency range (4.4–7.1 THz) is negative. This indicates that as heat flows from m40-4 to the interface at m40-1, a larger proportion of heat is transferred from high-frequency phonon modes to lower-frequency modes. This behavior arises because the cutoff frequency of m90 is approximately 5.5 THz, and heat carried by phonons with frequencies above 5.5 THz in m40 must shift to lower-frequency modes in m90.

Upon crossing into m90, the low-frequency part of $\Delta(\omega)$ transitions from negative values at m90-1 to zero at m90-4, while the high-frequency part decreases from positive values to zero. This further suggests a redistribution of heat from high-frequency to low-frequency phonons as the heat propagates through m90.

To quantify the rate at which the deviation decays as phonons travel away from the interface, we define the overall spectral deviation at a given location $x$ as:
\begin{equation}
\Delta(x) = \int |\Delta(\omega,x)| \, d\omega.
\label{eqn:delta}
\end{equation}
As shown in Fig.~\ref{fig:SI}d, $\Delta$ decreases monotonically with increasing distance $\left|x\right|$ from the m40-m90 interface ($x = 0$). The data are well-fitted by an exponential function of the form $c \cdot \exp(-\left|x\right|/\delta_{t})$, where $c$ is a constant and $\delta_{t}$ is a characteristic length that quantifies the distance required for $Q(\omega)$ to recover its bulk form. Specifically, to improve the accuracy of exponential fitting curves, we include 4 additional positions within each m40 or m90 material, resulting in a total of 8 data points for each fitting curve.

From curve fitting, the characteristic decay distances $\delta_{t}$ are determined to be 436~\AA\ for m40 and 247~\AA\ for m90. At these distances, the exponential function decays to 36.7\% of its peak value at $x = 0$. To reduce $\Delta$ to less than 5\% of its peak value—indicating the effective disappearance of nonequilibrium—the distance from the interface must reach approximately $3\delta_{t}$. This corresponds to distances of 1,308~\AA\ for m40 and 741~\AA\ for m90, where $Q(\omega)$ effectively returns to its bulk form.

This restoration of $Q(\omega)$ to its bulk form is due to anharmonic phonon scattering, which redistributes thermal energy among phonon modes of different frequencies. Consequently, we argue that $\delta_{t}$ is strongly correlated with the phonon MFP, as both are directly influenced by anharmonic scattering processes. Figure~\ref{fig:SI}e displays the MFPs of longitudinal acoustic (LA) and transverse acoustic (TA) phonons in m40 and m90, obtained through phonon SED analysis. The horizontal lines represent the $\delta_{t}$ values obtained in this study, which lie within the MFP ranges for m40 and m90. Notably, both the MFP and $\delta_{t}$ are longer in m40 than in m90, due to the higher phonon group velocities in m40, which arise from its lower atomic mass.

The evolution of $Q(\omega)$ along the heat flow direction provides direct evidence of nonequilibrium between phonon modes, which has been observed in various forms in prior studies. For example, previous works have reported significant temperature differences between phonon modes near interfaces \cite{zhou2013phonon,feng2017spectral,LI2023101063}. Furthermore, solutions to the Boltzmann transport equation in Refs.~\cite{LI2023101063,han2024PRB} have revealed nonequilibrium phonon distributions on both sides of an interface, leading to additional thermal resistance. In our recent study, we demonstrated substantial phonon nonequilibrium between coherent and incoherent phonons at the interface between heat baths and devices, which reduces the effective thermal conductivity of the device \cite{ma2022ex}. Unlike those previous studies, the heat flux spectra $Q(\omega)$ shown in Fig.~\ref{fig:SI}b provide direct evidence of heat transfer between phonon modes near the interface.

The vibrational density of states (vDOS), which characterizes the distribution of vibrational energy among phonon frequencies, is also examined. As shown in Fig.~\ref{fig:SI}f, the vDOS at all locations within the m40 and m90 materials closely resembles the vDOS of their respective bulk counterparts. This suggests that the vibrational energy distribution remains largely unchanged, even near the interface (e.g., at m40-1 and m90-1). This behavior contrasts sharply with the heat flux spectra $Q(\omega)$, which remain nearly the same from m40-1 to m90-1 in the vicinity of the interface, yet deviate significantly from their bulk counterparts.

\subsection{Double interfaces: Interlayer structures}

\begin{figure}
\centering 
\includegraphics[width=\textwidth]{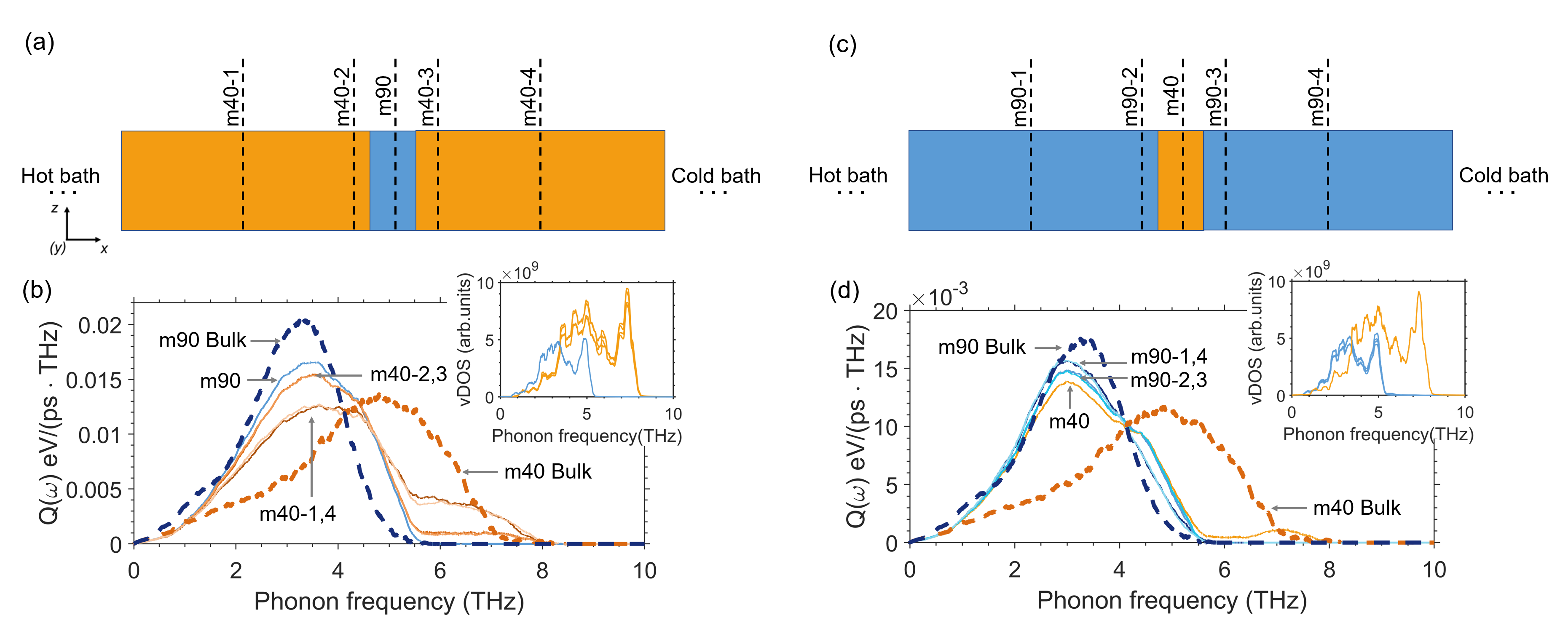}
\caption{Schematics of simulation setup: (a) the m40-m90-m40 sandwich structure, and (c) the m90-m40-m90 sandwich structure. Specifically, the 8-UC-thick interlayer is sandwiched between two 132-UC-thick contacts. Spectral heat flux $Q(\omega)$ for (b) the m40-m90-m40 sandwich structure and (d) the m90-m40-m90 sandwich structure, with the inlets showing the vDOS of the m40 (gold) and m90 (blue) layers. Additionally,  the solid curves refer to the $Q(\omega)$ at different positions, while the thick dashed curves represent the $Q(\omega)$ of bulk m40 and m90 materials.}
\label{fig:sandwich}
\end{figure}

The long distance (hundreds of angstroms) required for the heat flux spectra $Q(\omega)$ to return to equilibrium motivates further investigation into the case of double interfaces, where an 8-UC-thick ($\sim$42 Å) interlayer, much thinner than $\delta_{t}$, is sandwiched between two contacts made of a different material. As illustrated in Fig.~\ref{fig:sandwich}a, we first examine a thin m90 interlayer (8 UCs thick) placed between two large m40 contacts (each 132 UCs thick). Figure~\ref{fig:sandwich}b shows the heat flux spectra $Q(\omega)$ extracted from various locations, including m40-1, m40-2, m40-3, and m40-4 within the m40 contacts, as well as from m90 within the m90 interlayer. For comparison, the $Q(\omega)$ curves for bulk m40 and bulk m90, scaled to match the total heat flux of the other curves, are also shown. It is observed that within the m40 contacts, the heat flux spectra $Q(\omega)$ at m40-1/2/3/4 deviate significantly from the bulk m40 spectra, with deviations becoming more pronounced closer to the interlayer. This can be explained by the inability of high-frequency phonons in the m40 contacts, which are absent in m90, to propagate into the m90 interlayer. Consequently, the heat carried by these high-frequency phonons must first be transferred to lower-frequency modes that can be supported by the m90 interlayer. This process results in a nonequilibrium heat flux distribution across phonon modes, varying along the transport path across the interlayer.

A notable observation is that the $Q(\omega)$ spectra remain largely unchanged as heat flows from m40-2 (located within the left m40 contact, near the interlayer) into the m90 interlayer and then across into m40-3 (located within the right m40 contact, near the interlayer). These findings suggest that significant heat flux redistribution does not occur when phonons travel across the interlayer. Similar to the single-interface case shown in Fig.~\ref{fig:SI}c, heat flux redistribution primarily takes place over a distance of $\delta_{t}$ away from the interface; if the interlayer thickness is shorter than $\delta_{t}$, $Q(\omega)$ does not return to its bulk form.

We also examine the inverse case, where a thin m40 interlayer is sandwiched between two large m90 contacts, as depicted in Fig.~\ref{fig:sandwich}c. As shown in Fig.~\ref{fig:sandwich}d, the heat flux spectra for the m40 interlayer (m40) and the locations within the m90 contacts (m90-1/2/3/4) are quite similar, all deviating substantially from the bulk spectra of both m40 and m90. Notably, the high-frequency component ($\omega>5.5$ THz) of the bulk m40 heat flux spectra is significantly suppressed. This suppression occurs because the m90 material has a cutoff frequency of approximately 5.5 THz, as shown in Fig.~\ref{fig:SI}f. Consequently, almost no phonons with frequencies above 5.5 THz enter the m40 interlayer from the m90 contacts, resulting in minimal heat flux carried by phonons with frequencies above 5.5 THz within the interlayer.

The insets of Figs.~\ref{fig:sandwich}b and \ref{fig:sandwich}d display the vDOS at the locations where the heat flux spectra were calculated. In contrast to the behavior of the heat flux spectra, the vDOS of the interlayer remains nearly identical to that of its bulk material, with significant differences from the vDOS of the contact material. This observation is consistent with the single-interface cases discussed earlier (as shown in Fig.~\ref{fig:SI}f), indicating that the effect of the interface on vibrational properties is much weaker than its effect on the heat flux distribution $Q(\omega)$.

\begin{figure*}
\centering 
\includegraphics[width=0.95\textwidth]{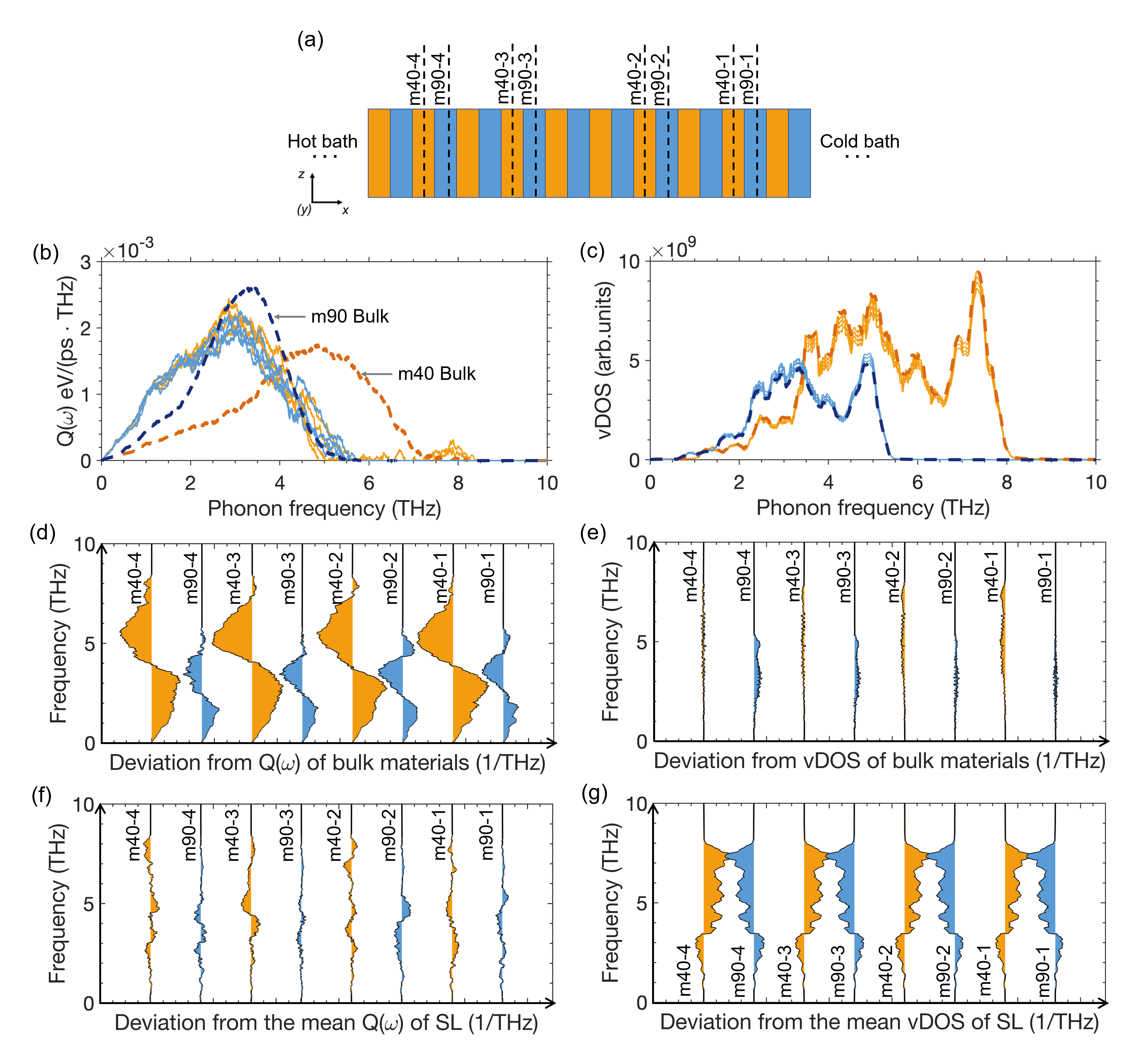}
\caption{Spectral heat flux $Q(\omega)$ and phonon vDOS of m40/m90 superlattice (NP64/d=8UC) at 30 K. (a) Schematic illustration of superlattice structure and positions selected to calculate $Q(\omega)$. (b) $Q(\omega)$ at different positions, where the solid curves refer to the $Q(\omega)$ of superlattice at different positions, while the thick dashed curves represent the $Q(\omega)$ of bulk materials. (c) vDOS at different positions, where the solid and dashed curves represent the vDOS of superlattice and bulk materials, respectively. (d) Deviation from $Q(\omega)$ of bulk materials. (e) Deviation from vDOS of bulk materials. (f) Deviation from $Q(\omega)$ of superlattice. (g) Deviation from vDOS of superlattice.} 
\label{fig:SL}
\end{figure*}

\subsection{Multiple interfaces: Superlattice structures}

Superlattices are unique structures characterized by periodic or aperiodic arrangements of multiple interfaces, exhibiting phonon behavior that differs significantly from that in bulk materials or single-interface structures \cite{landry2008complex,luckyanova2012coherent,wang2014decomposition,zhang2021coherent}. In this section, we investigate how $Q(\omega)$ evolves in superlattices where the material layers are thinner than or comparable to $\delta_{t}$.

The first superlattice structure we examine consists of 64 periods, with each period containing an 8-UC-thick m40 layer and an 8-UC-thick m90 layer. Here, we will denote this superlattice as NP64/d=8C, where the NP represents the number of periods. As illustrated in Fig.~\ref{fig:SL}a, we extract the heat flux spectra $Q(\omega)$ from eight locations: four within the middle of the m40 layers and four within the middle of the m90 layers. Figure~\ref{fig:SL}b shows that the $Q(\omega)$ curves in the superlattice are almost identical to each other, while deviating significantly from the bulk m40 and m90 spectra.

Figure~\ref{fig:SL}d presents the deviation ($\Delta(\omega)$, defined in Eq.~\ref{eqn:delta_omega}) of the local spectra from the corresponding bulk materials. The results reveal that the heat flux in the m40 layers of the superlattice is distinctly different from that in bulk m40, as evidenced by the significant gold-shaded regions in Fig.~\ref{fig:SL}d. A similar conclusion applies to the m90 layers, represented by the blue-shaded regions. However, when the average $Q(\omega)$ of all the m40 and m90 layers is used as the reference $Q(\omega)_{\text{ref}}$ in Eq.~\ref{eqn:delta_omega} (Fig.~\ref{fig:SL}f), the deviation becomes negligible. This indicates that the heat flux spectra remain nearly unchanged as heat flows from the hot bath to the cold bath across the entire superlattice, signifying minimal heat flux redistribution.

The absence of significant heat flux redistribution in the $d = 8$-UC superlattice is notable, as it suggests that heat flows through the structure as if it were a single material with no interfaces, eliminating the need for heat flux redistribution. This finding aligns with previous studies on the highly ballistic thermal transport properties of superlattices \cite{luckyanova2012coherent,wang2014decomposition}. Phonons with spatial coherence lengths and mean free paths greater than the superlattice period can interfere constructively to form new modes that travel ballistically through the superlattice without scattering at the interfaces. These ``coherent phonons'' follow the superlattice’s dispersion relation and dominate thermal transport. Our observations in Figs.~\ref{fig:SL}b and \ref{fig:SL}f confirm the importance of coherent phonons in governing thermal transport within superlattices.

Although the heat flux distribution $Q(\omega)$ is nearly identical in the m40 and m90 layers of the superlattice, vDOS remains distinctly different. As shown in Fig.~\ref{fig:SL}c, the vDOS for the m40 layers (gold curves) and the m90 layers (blue curves) corresponds to the bulk profiles of their respective materials. We quantify this distinction by calculating the deviation of the local vDOS in the m40 and m90 layers from their bulk counterparts using Eq.~\ref{eqn:delta_omega}. Figure~\ref{fig:SL}e shows that these deviations are minimal, in sharp contrast to the significant deviations observed in $Q(\omega)$ (Fig.~\ref{fig:SL}d). Furthermore, Fig.~\ref{fig:SL}g shows the deviation of the local vDOS from the average vDOS of all m40 and m90 layers in the superlattice, highlighting substantial differences. These results emphasize that, despite the similar heat flux distributions, the m40 and m90 layers retain their distinct vibrational properties, as reflected in their respective vDOS.

To better understand the contrasting behavior of $Q(\omega)$ and vDOS in the superlattice, it is essential to differentiate between these two quantities. $Q(\omega)$ represents the pathways for phonon energy transport, and the minimal deviation from the average $Q(\omega)$ in the superlattice suggests that coherent phonon transport, following the superlattice’s dispersion relation, is the dominant mechanism of thermal transport. In contrast, vDOS is not a direct measure of energy transport, though it can often qualitatively explain thermal transport behaviors in dissimilar materials \cite{english2012enhancing,wang2014phonon,chakraborty2017ultralow}. Instead, vDOS quantifies the available vibrational states at each energy level. Coherence effects related to energy transport do not significantly alter the vDOS. In fact, only a small fraction of vibrational modes in the m40 and m90 layers are coherent enough to form the coherent phonons that dominate heat transport in the superlattice. The majority of vibrational modes, which constitute most of the vDOS spectra, remain incoherent and localized within their respective material layers.

\begin{figure}
\centering 
\includegraphics[width=0.5\textwidth]{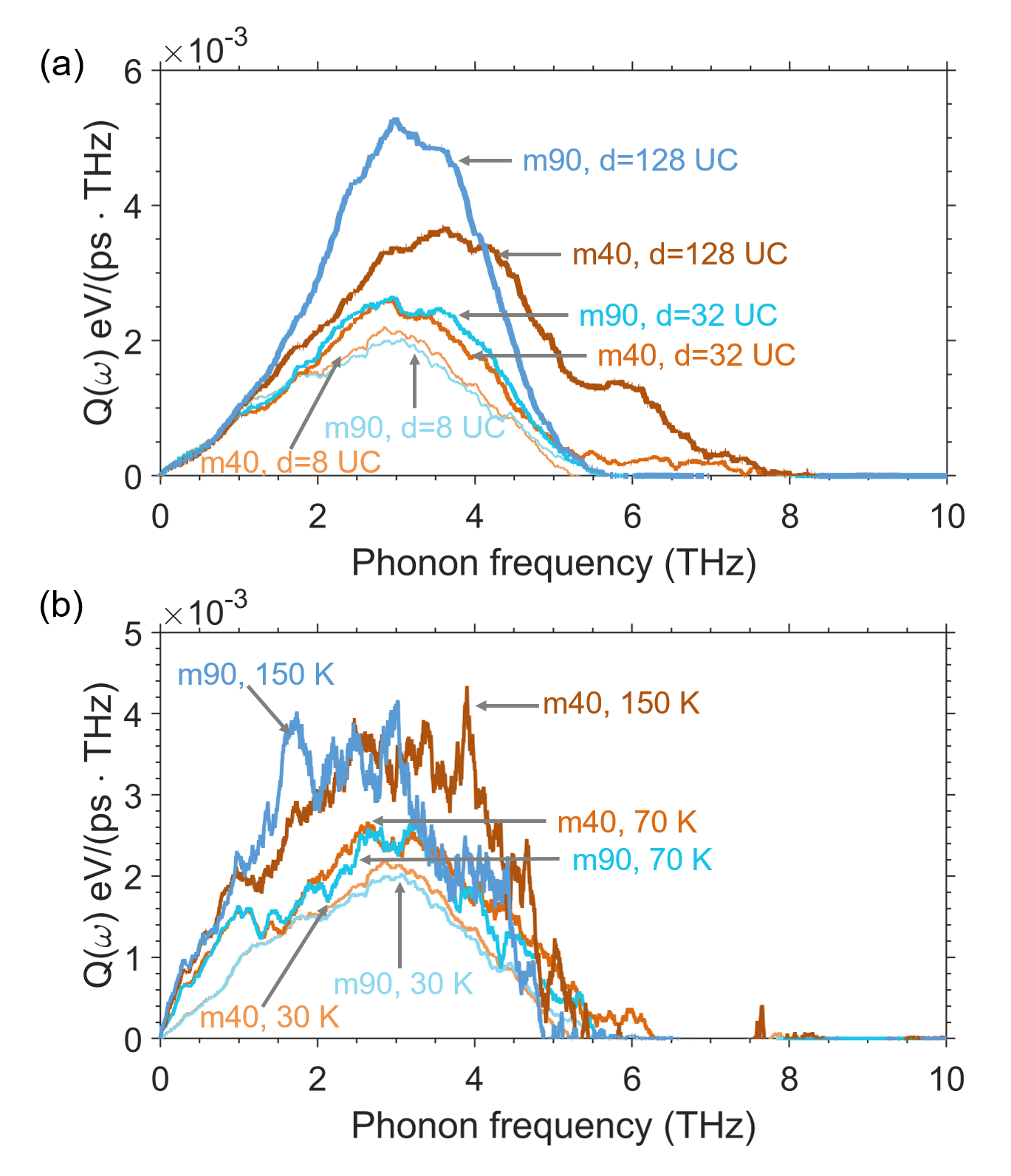}
\caption{(a) Period length dependent $Q(\omega)$ of various superlattices with increasing period length (d=8UC, d=32 UC, and d=128 UC) at 30 K. (b) Temperature dependent $Q(\omega)$ of NP64/d=8UC superlattices at 30 K, 70 K, and 150 K. }
\label{fig:Q_vs_d_T}
\end{figure}

To further clarify the influence of $\delta_{t}$ on heat flux redistribution, we investigate scenarios where the layer thickness $d$ of the superlattice approaches $\delta_{t}$. This can be accomplished by either increasing $d$ or reducing $\delta_{t}$. However, it is important to note that the resolution of extracting $Q(\omega)$ from molecular dynamics simulations limits our ability to obtain high-quality $Q(\omega)$ spectra when $d$ becomes significantly larger than $\delta_{t}$. In such cases, the heat flux data often exhibit excessive noise, hindering quantitative analysis.

First, we examine m40-m90 superlattices with thicker layers, specifically at $d = 32$ UC and $d = 128$ UC. As shown in Fig.~\ref{fig:Q_vs_d_T}a, the $Q(\omega)$ curves for the $d = 32$-UC superlattice remain similar across all m40 and m90 layers, akin to the behavior observed in the $d = 8$-UC case in Fig.~\ref{fig:SL}. In contrast, for the $d = 128$-UC superlattice, the $Q(\omega)$ curves of the m40 and m90 layers diverge considerably, each reverting to the $Q(\omega)$ profiles of their respective bulk forms.

Next, we explore cases where $\delta_{t}$ is reduced, noting that $\delta_{t}$ decreases at higher temperatures due to increased anharmonic phonon scattering. As illustrated in Fig.~\ref{fig:Q_vs_d_T}b, the $Q(\omega)$ of the m40 and m90 layers become more distinct at elevated temperatures. At 30 K and 70 K, the m40 and m90 curves remain nearly indistinguishable, but at 150 K, the m90 curve shifts to lower frequencies while the m40 curve shifts to higher frequencies. This suggests that at higher temperatures, the heat flux distribution thermalizes more strongly, approaching the respective bulk distributions.

\section{Conclusion}
In conclusion, our comprehensive analysis of phonon spectral heat flux reveals key insights into heat transport across interfaces and within superlattice structures. When phonons traverse a single interface between two dissimilar materials with distinct phonon spectra, we observe a redistribution of heat flux among different phonon modes. Importantly, this redistribution does not occur instantaneously upon crossing the interface, suggesting that phonon transmission is predominantly elastic at the interface. Instead, the redistribution unfolds over a characteristic distance, which we estimate to be on the order of hundreds of angstroms, based on the phonon mean free paths. For the conceptual materials m40 and m90, this characteristic thermalization length, $\delta_{t}$, was extracted through curve fitting, but the approach is extendable to practical materials systems, such as silicon/germanium, providing a method to quantify $\delta_{t}$ for real-world applications.

In superlattices composed of alternating thin layers of m40 and m90, where the layer thickness is comparable to or smaller than $\delta_{t}$, we observe nearly identical heat flux distributions, $Q(\omega)$, in both materials, in stark contrast to the $Q(\omega)$ of their respective bulk counterparts. This behavior underscores the role of coherent phonons, which dominate heat transport in these superlattices, maintaining the same heat flux distribution across the layers. Despite this, the vDOS within the layers remains consistent with that of their bulk forms, indicating that the vibrational properties of each material are preserved even as their heat flux behavior converges.

Finally, we demonstrated that as the layer thickness of the superlattices approaches $\delta_{t}$, the heat flux distributions in the two materials begin to diverge once again. This divergence results from anharmonic phonon scattering within the thicker layers, which redistributes thermal energy among phonon modes, thereby reverting the heat flux distribution $Q(\omega)$ towards its bulk form. These findings provide a deeper understanding of the interplay between phonon coherence, thermalization length, and anharmonic scattering in complex material systems, and offer valuable insights for the design of nanostructured materials for thermal management applications.

\clearpage
\begin{acknowledgments}
The authors gratefully acknowledge the financial support from the National Science Foundation (CBET-2047109). Cui and Wang also extend their thanks to the National Science Foundation EPSCoR Research Infrastructure Program (OIA-2033424). Ma was also partially supported by the National Science Foundation (Award No. 1953300). Additionally, the authors would like to acknowledge the support provided by the Research and Innovation team and the Cyberinfrastructure Team in the Office of Information Technology at the University of Nevada, Reno, for facilitating access to the Pronghorn High-Performance Computing Cluster.
\end{acknowledgments}

\section*{DATA AVAILABILITY}
The data that support the findings of this study are available from the corresponding author upon reasonable request.

\bibliography{references}

\end{document}